\documentstyle[twoside,fleqn,espcrc2,epsfig,here]{article}

\def\beq{\begin{equation}}
\def\eeq{\end{equation}}
\def\MSbar {\hbox{$\overline{\hbox{\tiny MS}}\,$}}
\def\eff{\hbox{\tiny eff}}

\def\PV{\hbox{\tiny PV}}

\def\BLM{\hbox{\tiny BLM}}
\def\NLO{\hbox{\tiny NLO}}

\def\1loop{\hbox{\tiny 1-loop}}
\def\2loop{\hbox{\tiny 2-loop}}

\newcommand{\AmS}{{\protect\the\textfont2
  A\kern-.1667em\lower.5ex\hbox{M}\kern-.125emS}}

\title{Power corrections and renormalon resummation for the average
thrust\thanks{Talk given by 
E. Gardi at the QCD'99 conference, Montpellier, July 1999.}}

\author{Einan Gardi \,\,\,and\,\,\, Georges Grunberg
\address{
Centre de Physique Th\'eorique de l'Ecole Polytechnique
91128 Palaiseau Cedex, France
\\email: {\tt \,\,gardi@cpht.polytechnique.fr,\,\,
grunberg@cpht.polytechnique.fr}}
}

\begin{document}

\begin{abstract}
Infrared power corrections for the average thrust $\left< T \right>$ in 
$e^+e^-$ annihilation are analyzed in the framework of renormalon 
resummation, motivated by analogy with the skeleton expansion in QED 
and the BLM approach. Performing the ``massive gluon'' renormalon 
integral a renormalization scheme invariant result is obtained. 
We find that a major part of the discrepancy between the known 
next-to-leading order (NLO) calculation and experiment can be
explained by resummation of higher order perturbative terms. This fact
does not preclude the infrared finite coupling interpretation with 
a substantial $1/Q$ power term. Fitting the regularized perturbative 
sum plus a $1/Q$ term to experimental data yields 
\hbox{$\alpha_s^{\hbox{$\overline{\hbox{\tiny MS}}\,$}}({\rm M_Z})
=0.110\pm  0.006$}.
\end{abstract}

\maketitle

\section{Introduction}

Power corrections to event shape observables in $e^+e^-$ annihilation 
have been an active field of research in the recent years. 
Event shapes, as opposed to other inclusive observables, 
do not have an operator product expansion, so there is no 
established field theoretic framework to analyse them beyond the
perturbative level. On the other hand, the experimental data 
now available cover a wide range of scales and thus could provide an
opportunity to test QCD and extract a precise value of~$\alpha_s$.  

The state of the art in perturbative calculations of average event shape
variables is ${\cal O}(\alpha_s^2)$, i.e. NLO. 
It turns out that experimental data are not well described by 
these perturbative expressions, unless explicit power corrections, that
may be associated with hadronization, are introduced. 
Renormalon phenomenology allows to predict the form of the power terms
while their magnitude is determined by experimental fits. 
 
In the work reported here \cite{ours} we assume the existence of an
Abelian like ``dressed skeleton expansion'' in QCD and calculate the 
single dressed gluon contribution using the dispersive approach. 
This way we perform at once all order resummation of perturbative terms 
which are related to the running coupling (renormalons) and 
parametrization of power corrections. 
We discuss the ambiguity between the perturbative sum and the power 
corrections and show that the resummation is essential in order to
extract the correct value of $\alpha_s$ from experimental data.

\section{Average thrust in perturbation theory}

To demonstrate the method proposed we concentrate on one specific 
observable, the average thrust, defined by
\beq
T=\frac{\sum_i \left\vert \vec{p}_i \cdot \vec{n}_T\right\vert}
{\sum _i \left\vert \vec{p}_i \right\vert},
\label{T_def}
\eeq
where $i$ runs over all the particles in the final state,
$\vec{p}_i$ are the 3-momenta of the particles and $\vec{n}_T$ is the
thrust axis which is set such that $T$ is
maximized. It is useful to define $t\equiv 1-T$ which vanishes in the
2-jet limit.

\begin{figure}[htb]
\mbox{\kern-0.1cm
\epsfig{file=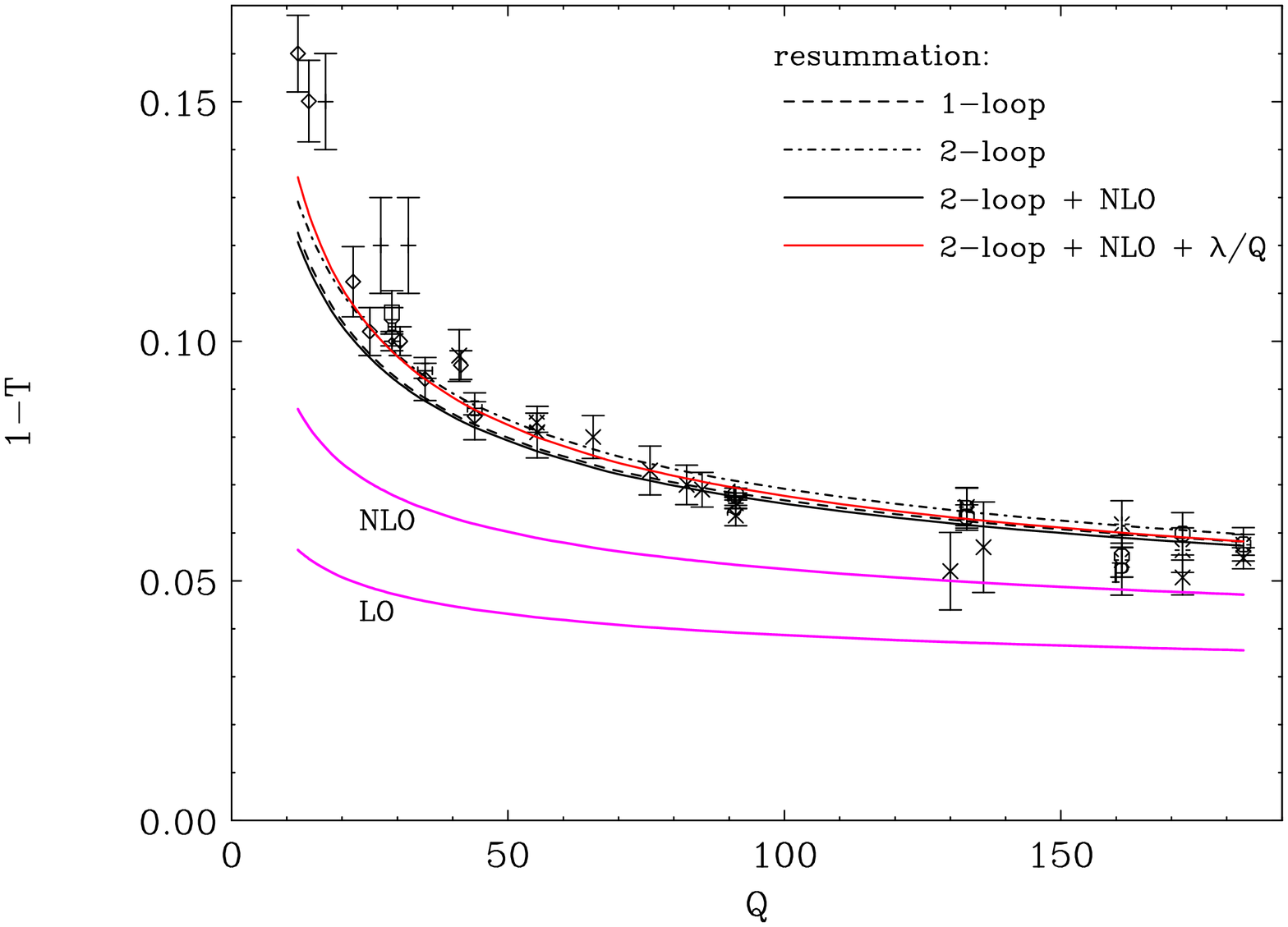,
width=5.5truecm,angle=90}
}
\vspace{-35pt}
\caption{$\left<t\right>$ as a function of $Q=\sqrt{s}$ for
$\alpha_s^{\MSbar}(M_Z)=0.117$. Naive
$\overline{\rm MS}$ results (LO, NLO) and resummation results vs
experimental data. }
\label{K_117_PV}
\end{figure}

Being collinear and infrared safe, the average thrust can be
calculated in perturbative QCD to yield
\beq 
\left<t \right>_{\NLO}(Q^2) = \frac{C_F}{2}\,
\left[t_0 \,a_{\MSbar}(Q^2) \,+\, t_1 \,a_{\MSbar}^2(Q^2)\right]
\label{t_MSbar}
\eeq
where $a=\alpha_s/\pi$, $C_F=\frac{4}{3}$ and the coefficients are 
$t_0 = 1.58$ and $t_1 = 23.7-1.69\, N_f$ (see refs. in \cite{ours}). 
Using the world average value of $\alpha_s$, $\alpha_s^{\MSbar}({\rm
M_Z})=0.117$, the NLO perturbative result (\ref{t_MSbar}) turns out to
be quite far from experimental data. This is shown in fig.~\ref{K_117_PV}.
From the figure it is clear that the NLO 
correction is quite significant, so the perturbative series 
truncated at this order is not very reliable. 
Furthermore, the renormalization scale dependence is
significant. It follows that higher order corrections that are
related to the running coupling cannot be ignored.

\section{Renormalons and power corrections}

In order to take running coupling effects into account it is useful \cite{BLM}
to assume, in analogy with the Abelian theory, that there exists
a ``dressed skeleton expansion''. Then, the most important corrections
which correspond to a single dressed gluon can be written in the form
of a renormalon integral
\beq
D_0=\int_0^{\infty}\frac{dk^2}{k^2}\,\phi(k^2/Q^2)\,\bar{a}(k^2)
\label{D0}
\eeq
where $\bar{a}$ represents a specific ``skeleton effective charge'', not yet
determined in QCD (in \cite{ours} several schemes were used).
As opposed to the standard perturbative approch (\ref{t_MSbar}),
by performing the integral (\ref{D0}) over all scales one avoids completely
renormalization scale dependence.
 
The integral (\ref{D0}) represents a non Borel summable power series. 
Indeed, it involves integration over the coupling
constant in the infrared which is ill-defined in perturbation theory
due to Landau singularities. The ambiguous integral (\ref{D0}) can be defined 
mathematically, e.g. as a principle value of the Borel sum:
$D_0\vert_{\PV}$. This, however, does not solve the physical problem 
of infrared renormalons: 
information on large distances is required to fix the ambiguity.
The perturbative calculation contains some information about the ambiguity:
if the leading term in the small momentum expansion of $\phi$ 
is \hbox{$\phi(k^2/Q^2)=C_n(k^2/Q^2)^n$}, the leading infrared
renormalon is located at $n/\beta_0$ in the Borel plane, and a power 
correction of the form $1/Q^{2n}$ is expected. 
Having no way to handle the problem on the non-perturbative level, it
is natural to attempt a fit of the form $D_0\vert_{\PV}+\lambda/Q^{2n}$.

A stronger assumption \cite{DMW} is that the ``skeleton coupling'' can
be defined on the non-perturbative level down to the infrared. Then
the integral (\ref{D0}) should give at once the perturbative 
result plus the correct power term. Since the infrared coupling is not
known, it is considered as a non-perturbative parameter. Using 
the cutoff regularization of (\ref{D0}), one fits the data with 
\hbox{$D_0\vert_{\mu_I}+\lambda_{\mu_I}/Q^{2n}$}, where
\beq
D_0\vert_{\mu_I}=\int_{\mu_I^2}^{\infty}\frac
{dk^2}{k^2}\,\phi(k^2/Q^2)\,\bar{a}(k^2)
\label{D0_mu_I}
\eeq
is fully under control in perturbation theory and the normalization of
the power term
\beq
\lambda_{\mu_I}=C_n\int_0^{\mu_I^2}\frac{dk^2}{k^2}\,k^{2n}\,\bar{a}(k^2) 
\label{moment}
\eeq
is a perturbatively calculable coefficient times a moment of the
coupling in the infrared. Since the coupling is assumed to be
universal, the magnitude of power corrections can be compared between
different observables \cite{DMW}.
 
The generalization of this approach to Minkowskian observables such as
the thrust was discussed in \cite{DMW,ours}. At the level
of a single gluon emission it is based on a ``gluon mass'' renormalon 
integral \cite{BB}
\begin{eqnarray}
\label{Rapt} 
R_{0} &\equiv&
\int_0^\infty{d\mu^2\over\mu^2}\ \bar{a}_{\eff}(\mu^2)\ 
\dot{{\cal F}}(\mu^2/Q^2) 
\end{eqnarray}
where the characteristic function ${\cal F}$ is calculated based on
the matrix element for the emission of one massive gluon and
$\bar{a}_{\eff}$ is related to the time-like discontinuity of the coupling.
Regularizations of the perturbative sum (\ref{Rapt}) with $\bar{a}$ at
one or two loops were discussed in \cite{ours}. We skip it here for brevity.

\section{Fitting experimental data}

In order to perform renormalon resummation at the level of a single gluon
emission, we calculated the characteristic function for the 
thrust~\cite{ours}. Then $R_0$ is evaluated taking either the one or two loop
coupling. Since $R_0$ does not exhaust the full NLO correction of
eq.~(\ref{t_MSbar}), we add an explicit NLO correction,
$\delta_{\NLO}=\left(t_1-t_1^0\right)a_{\MSbar}^2$, where $t_1$
corresponds to (\ref{t_MSbar}) and $t_1^0$ is the piece included in $R_0$.
It was shown in \cite{ours} that the Abelian part of $t_1^0$ almost
coincides with that of $t_1$ in spite of the non-inclusive nature of
the thrust. This is crucial for the applicability of the current
resummation approach. Finally, we add an explicit power correction 
$\lambda/Q^{2n}=\lambda/Q$ where the power $n=\frac12$ 
is determined from the small $\mu^2$ expansion of ${\cal F}$.
Thus, using the PV regularization of $R_0$, we fit the data with
\beq
\left<t\right>=\frac{C_F}{2}\left[R_0\vert_{\PV}+\delta_{\NLO}\right]
+\frac{\lambda}{Q}.
\label{inc_lambda}
\eeq
The resummation as well as the fit results are presented in
fig.~\ref{K_117_PV} for the world average value of $\alpha_s$.
The resummation by itself is quite close to the data.
Note also the stability of the result as $\bar{a}$ is promoted from
one to two loops.
Next, we fit also the value of $\alpha_s$. The best fit 
$\chi^2/{\rm point}=1.35$, obtained with
$\alpha_s^{\MSbar}(M_Z)=0.110\pm 0.006$ and $\lambda=0.62\pm 0.11$, 
is shown in fig.~\ref{PV_cutoff_110}. 

\section{Truncated expansions in $\overline{\rm MS}$}

\begin{figure}[htb]
\mbox{\kern-0.1cm
\epsfig{file=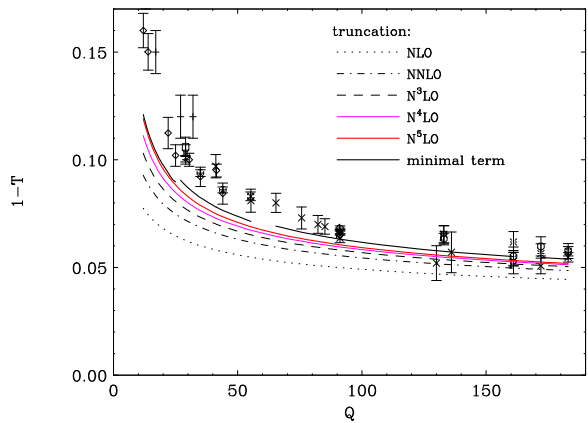,width=5.5truecm,angle=90}
}
\vspace{-35pt}
\caption{$\left<t\right>_{\rm pert}$ as a function of $Q$ for 
\hbox{$\alpha_s^{\hbox{$\overline{\hbox{\tiny MS}}\,$}}({\rm M_Z})=0.111$}. The
perturbative series is gradually improved by adding terms from $R_0$.} 
\label{pert}
\end{figure}

It is interesting to compare the resummation to a truncated
expansion in $a_{\MSbar}$. Expanding $R_0$ with $\bar{a}$ at 1-loop 
we obtain a series of the
form $\sum_{i=0}^{\infty} t_i^0 a_{\MSbar}^{i+1}$. The coefficients
$t_i^0$ can be considered as predictions for the perturbative
coefficients $t_i$, provided the coupling $\bar{a}$ is close to the correct
``skeleton coupling'', and provided that the ``leading skeleton''
$R_0$ is indeed dominant already in the sub-asymptotic regime. 
The significance of the NNLO and further sub-leading corrections which
correspond to the dissociation of the emitted gluon is demonstrated in
fig.~\ref{pert}. The figure shows 
\begin{eqnarray}
\label{trunc_fit}
\left<t\right>_{\rm pert}&\equiv&\left<t\right>_{\NLO}+
\frac{C_F}{2} \sum_{i=3}^{k}\,t_{i-1}^0\,a_{\MSbar}^{i}(Q^2) 
\end{eqnarray}
where $k$ is the order of truncation for $k=2$ through $6$. At these orders
the series is still convergent.
We conclude that a major part of the discrepancy between the NLO
result and experiment is due to neglecting this particular class of higher
order corrections. 
Next consider a fit to experimental data based on the truncated 
expansion of $R_0$ (\ref{trunc_fit}): 
\hbox{$\left<t\right>_{\rm pert}+{\lambda_{\rm pert}}/{Q}$}.
The fit results are listed in table~\ref{trunc_fit_tab}. 
\begin{table}[H]
\vspace{-20pt}
\caption{Fit results based on the truncated expansion~in~$\overline{\rm MS}$}
\vspace{5pt}
\label{trunc_fit_tab}
\[
\begin{array}{ccc}
\hline
\,k\,&\,\alpha_s^{\MSbar}({\rm M_Z})\,& \,\lambda_{\rm pert}\, ({\rm GeV})\,\\
\hline
2&0.128&0.72\\
3&0.118&0.65\\
4&0.115&0.58\\
5&0.114&0.50\\
6&0.114&0.40\\
\hline
{\rm PV}&0.111&0.73\\
\hline
\end{array}
\]
\vspace{-20pt}
\end{table}
\noindent
The quality of the fit is roughly the same in all cases.
We see that the resummation is absolutely necessary in order to extract
a reliable value~of~$\alpha_s$. 

\section{Infrared cutoff regularization}

\begin{figure}[htb]
\mbox{\kern-0.1cm
\epsfig{file=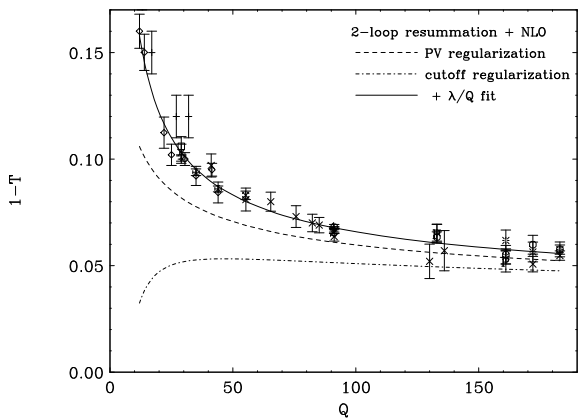,width=5.5truecm,angle=90}
}
\vspace{-35pt}
\caption{$\left<t\right>$ as a function of $Q$ for
\hbox{$\alpha_s^{\hbox{$\overline{\hbox{\tiny MS}}\,$}}({\rm M_Z})=0.110$}.
Resummation in different regularizations and a fitted curve based on
either of them.}
\label{PV_cutoff_110}
\end{figure}

Putting an infrared cutoff $\mu_I$ on the space-like momentum
(\ref{D0_mu_I}) we separate at once the perturbative and non-perturbative
regimes as well as large and small distances. The cutoff regularized
sum, $R_0\vert_{\mu_I}$ for
\hbox{$\mu_I=2\,{\rm GeV}$} is shown in fig.~\ref{PV_cutoff_110} together
with the PV regularization. 
For the lower $Q$ values the difference between the two regularizations
is large. This means that a large contribution to $R_0\vert_{\PV}$
comes from momentum scales below $2\,{\rm GeV}$, where the coupling is not 
controlled by perturbation theory. Finally fitting the data 
\beq
\left<t\right>=\frac{C_F}{2}\left[R_0\vert_{\mu_i}+\delta_{\NLO}\right]
+\frac{\lambda_{\mu_I}}{Q}
\label{inc_lambda_cutoff}
\eeq
we get the same result as with the PV regularization
(\ref{inc_lambda}). The only difference is in the required power
correction $\lambda_{\mu_I}=1.49$.
This is general: the two regularizations differ 
just by (calculable) infrared power corrections -- in our case $1/Q$
-- and are therefore equivalent once the appropriate power terms are included.

\section{Conclusions}

The assumption of a ``skeleton expansion'' implies that resummation 
of perturbation theory and parametrization of power corrections must
be performed together.
For the thrust renormalon resummation is significant and closes 
most of the gap between the standard perturbative 
result (NLO) and experiment. The resummation is crucial to extract 
the correct value of $\alpha_s$.

The infrared sensitivity of the thrust leads to ambiguity in
the resummation, which is settled by fitting a $1/Q$ term.
In the infrared cutoff regularization this power term is substantial.

The resummation leads to a renormalization scale invariant
result. In the BLM approach it corresponds to a low
renormalization point in $\overline{\rm MS}$, 
\hbox{$\mu_{\BLM}^{\MSbar}\simeq 0.0447\,Q$}.

\vspace{14pt}


{\bf E. De Rafael} {\it What is the physical meaning, in QCD, of the
scale of the $1/Q$ correction? What other processes are sensitive to
these corrections?}\\

{\it  
The first question is basically an open one. Deeper understanding 
could hopefully be gained once renormalon 
phenomenology is supported by more rigorous field theoretic methods.
In the infrared finite coupling approach the $1/Q$ correction 
is understood as a moment (\ref{moment}) of a universal infrared 
finite coupling. 
This allows comparison between observables -- for 
example the C parameter is sensitive to similar $1/Q$ corrections. 
In the framework of shape-functions~\cite{shape-functions}, a relation 
with the energy-momentum tensor was suggested.
}

\end{document}